\documentclass{article}
\usepackage{spconf,amsmath,graphicx}

\makeatletter
\def\blfootnote{\gdef\@thefnmark{}\@footnotetext}
\makeatother

\title{Towards achieving robust universal neural vocoding}
\name{\parbox{\linewidth}{\centering
Jaime Lorenzo-Trueba, Thomas Drugman, Javier Latorre, Thomas Merritt, Bartosz Putrycz, \\
Roberto Barra-Chicote, Alexis Moinet, Vatsal Aggarwal}
}
\address{Amazon.com}

\begin{document}
%
\maketitle
\begin{abstract}
This paper explores the potential universality of neural vocoders. We train a WaveRNN-based vocoder on 74 speakers coming from 17 languages. This vocoder is shown to be capable of generating speech of consistently good quality (98\% relative mean MUSHRA when compared to natural speech) regardless of whether the input spectrogram comes from a speaker or style seen during training or from an out-of-domain scenario when the recording conditions are studio-quality. When the recordings show significant changes in quality, or when moving towards non-speech vocalizations or singing, the vocoder still significantly outperforms speaker-dependent vocoders, but operates at a lower average relative MUSHRA of 75\%. These results are shown to be consistent across languages, regardless of them being seen during training (e.g. English or Japanese) or unseen (e.g. Wolof, Swahili, Ahmaric).

\end{abstract}
\vspace{-2mm}
\section{Introduction}
\label{sec:intro}
\vspace{-1mm}

\blfootnote{Corresponding author email: truebaj@amazon.com. Paper accepted on Interspeech 2019.}

Statistical parametric speech synthesis (SPSS) has seen a paradigm change recently, mainly thanks to the introduction of a number of autoregressive models \cite{van2016wavenet, kalchbrenner2018efficient, shen2017natural, ping2018, merritt2018comprehensive,prenger2018waveglow}, turning into what can be termed statistical speech waveform synthesis (SSWS) \cite{merritt2018comprehensive}. This change has closed the gap in naturalness between statistical text to speech (TTS) and natural recordings whilst maintaining the flexibility of statistical models.

In the case of traditional vocoding \cite{kawahara2001aperiodicity, morise2016WORLD, drugman2012deterministic, Macon:1996}, approaches commonly relied on simplified models (e.g. source-filter model \cite{fant1985four}) that were defined by acoustic features such as 
voicing decisions, the fundamental frequency (F0), mel-generalized cepstrum (MGC) or band aperiodicities. The quality of those traditional vocoders was limited by the assumptions made by the underlying model and the difficulty to accurately estimate the features from the speech signal \cite{merritt2014investigating,merritt2015attributing}. 

Traditional waveform generation algorithms, while capable of generating speech from their spectral representation such as Griffin-Lim \cite{griffin1984signal}, are not capable of generating speech with acceptable naturalness. This is due to the lack of phase information in the short-time Fourier transform (STFT).

Neural vocoders are a data-driven method where neural networks learn to reconstruct an audio waveform from acoustic features \cite{van2016wavenet, kalchbrenner2018efficient, jin2018fftnet,prenger2018waveglow}. They allow us to overcome the shortcomings of traditional methods \cite{wang2018comparison} at a very significant cost in computation power and data requirements.  However, due to sparsity (it is unlikely that we will ever be able to cover all possible human-generated sounds in the training data) the neural vocoder models are prone to over-fit to the training speaker characteristics and have poor generalization capabilities \cite{arik2019fast}. 
Several recent studies attempted to improve the adaptation capabilities of such models \cite{wu2018rapid, sisman2018voice}, commonly using explicit speaker information (either as a onehot encoding or some other form of speaker embedding) \cite{liu2018wavenet}. There are however reports in literature of initial successes training neural vocoders without providing explicit speaker information \cite{hayashi2017investigation,jia2018transfer}, however the investigation either did not provide significant improvements in terms of robustness or did not cover the details on how the model handles changes in domain or unseen speakers.

This contributions of this paper are: 1) we demonstrate that a speaker encoding is not required to train a high-quality Speaker-Independent (SI) WaveRNN-based \cite{kalchbrenner2018efficient} neural vocoder; 2) our SI neural vocoder can effectively synthesise speakers that were unseen during training, which is not possible with vocoders trained with explicit speaker information or with a speaker-dependent approach; 3) we study the robustness and potential universality of our SI neural vocoder on a large diversity of unseen conditions (e.g. language, phonation, noise or speaking style).

\vspace{-5mm}

\section{System description}
\label{sec:system}\label{sec:model}
\vspace{-2mm}
Even though CNN-based systems have been thoroughly researched and real-time implementations have been proposed \cite{ping2018, oord2017parallel}, it is known that they are prone to instabilities \cite{wu2018collapsed} which occasionally affect perceptual quality. RNN-based systems, on the other hand, can be expected to provide a more stable output due to the persistence of the hidden state, at least when vocoding, in which context is not critical beyond the closest spectrograms (a known characteristic of RNNs).

\begin{figure}[tb!]
	\centering
	\includegraphics[width=\linewidth]{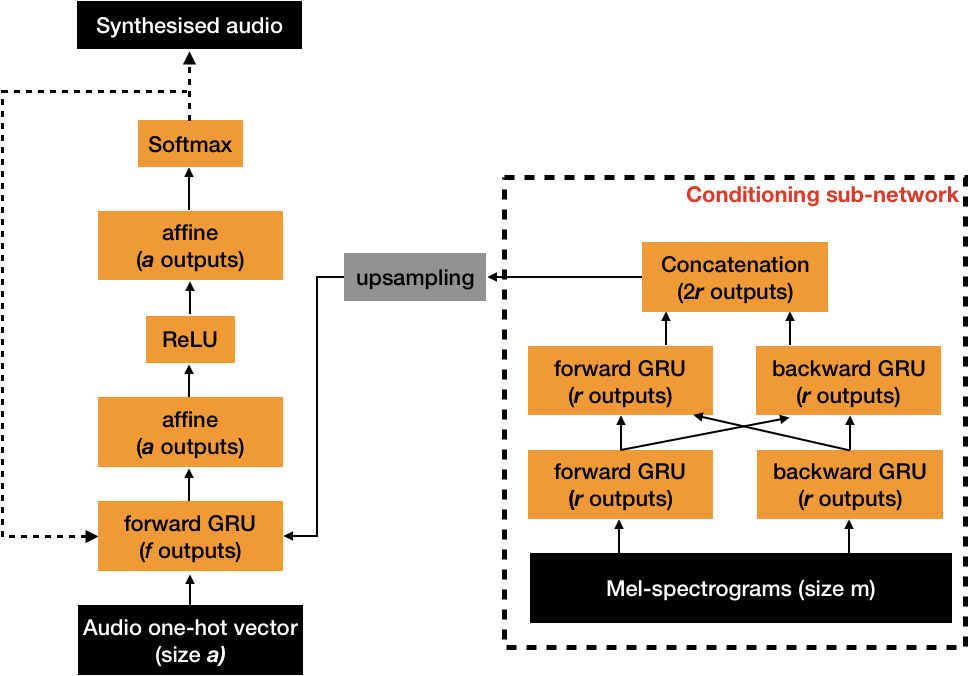}
	\vspace{-5mm}
	\caption{Block diagram of system RNN\_MS}
	\label{fig:rnn-neural}
	\vspace{-5mm}
\end{figure}

The structure of the neural vocoder system used in this paper (heavily inspired by WaveRNN \cite{kalchbrenner2018efficient}, only with minor changes in the conditioning network) is described in Figure \ref{fig:rnn-neural}. We refer to this system as RNN\_MS. The autoregressive side consists of a single forward GRU (hidden size of 896) and a pair of affine layers followed by a softmax layer with 1024 outputs, predicting the 10-bit mu-law samples for a 24 kHz sampling rate. The conditioning network consists of a pair of bi-directional gated recurrent units (GRUs) with a hidden size of 128. The mel-spectrograms used for conditioning the network were extracted using Librosa library \cite{mcfee2015librosa}, with 80 coefficients and frequencies ranging from 50 Hz to 12 kHz.

We trained system RNN\_MS in 4 different configurations, whose details are shown in Table \ref{tab:training}. First three SD systems were trained on American English speakers, two female (F1 \& F2) and 1 male (M1) from our internal corpora.

We also trained 3 multi-speaker vocoders, one with all the training data from the 3 SD voices ($3Spk$), another one with 7 American English speakers ($7Spk$) comprising 4 females, 2 males and 1 child but with restricted amounts of training data per speaker (5000 utterances). This $7Spk$ neural vocoder aims to check whether variability or data (i.e. $3Spk$) are more important for robustness in general. Finally we trained what is introduced as our universal neural vocoder with 74 different voices, 22 males and 52 females, extracted from 17 languages, with approx. 2000 utterances per speaker. This neural vocoder was designed with the expectation of being generalizable to any incoming speaker regardless of whether it was seen during training or not.


\begin{table}[tb!]
	\caption{Summary of the training data of the different RNN-based vocoders.}
	\footnotesize
	\centering
	\label{tab:training}
	\begin{tabular}{|c|c|l|l|}
		\hline
		\multicolumn{1}{|l|}{\textbf{Vocoder}} & \multicolumn{1}{l|}{\textbf{Speakers}} & \textbf{Utterances} & \textbf{Language} \\ \hline
		\textbf{F1 (SD)} & 1 & 22000 & US English \\ \hline
		\textbf{F2 (SD)} & 1 & 15000 & US English \\ \hline
		\textbf{M1 (SD)} & 1 & 15000 & US English \\ \hline
		\textbf{3spk} & 3 & 52000 & US English \\ \hline
		\textbf{7spk} & 7 & 35000 & US English \\ \hline
		\textbf{Univ} & 74 & 149134 & Multiple (17) \\ \hline
	\end{tabular}
	\vspace{-5mm}
\end{table}

\vspace{-3mm}
\section{Experimental protocol}
\label{sec:eval}
 \vspace{-2mm}
 To properly characterize the generalization capabilities of the different vocoders in terms of naturalness we considered a number of scenarios, but always considering oracle spectrograms directly extracted from recordings. 
 First of all a topline scenario in which we generated speech from speakers present in the training data of all the vocoders, but with utterances not seen during training (section \ref{sec:ind}). 
 Then, we also generated speech in scenarios partially out-of-domain from the training data: a mixture of male and female neutral speakers extracted from VCTK \cite{vctk} for English or from the NITech Japanese samples database \cite{zen2007hmm}. We also considered audiobook speech extracted from Blizzard2016 development set \cite{king2016blizzard}, which was out of domain in terms of speaker, speaking style but as in all previous cases, recorded with studio-quality.
 
Finally we considered a number of out-of-domain scenarios ranging from: i) different voice qualities \cite{airas2007comparison}, ii) irregular recording conditions (i.e. background noise \cite{valentini2017noisy}, reverberation \cite{valentini2016reverberant}, or both \cite{valentini2017noisyrev}), iii) unseen languages (Ahmaric, Swahili and Wolof) recorded in sub-optimal recording situations \cite{gauthier2016collecting} (i.e. significant reverberation, or poor quality audio), iv) singing extracted from publicly available music corpora \cite{SiSEC16}, v) non-speech vocalizations \cite{lima2013voices}.

The naturalness perceptual evaluation was designed as a MUltiple Stimuli with Hidden Reference and Anchor (MUSHRA) test \cite{recommendation2001bs}, where the participants were presented with the systems being evaluated side-by-side, asked to rate them in terms of naturalness from 0 (very poor) to 100 (completely natural), but modified so as not to force at least one 100 rated system. The test consisted of 200 randomly-selected utterances, not included in the training data. Evaluations were conducted with self-reported native American English speakers using Amazon Mechanical Turk. 50 listeners participated in each evaluation, balanced so that every utterance was rated by 5 listeners, each rating 20 screens.

Paired Student T-tests with Holm-Bonferroni correction were used to validate the statistical significance of the differences between systems, considering it validated when $p-value < 0.01$. We use the ratio between the mean MUSHRA score of a system and natural speech, we refer to this as 'relative MUSHRA', to illustrate the gap with the reference. 

\vspace{-3mm}
\section{Results}
\label{sec:res}

\vspace{-2mm}
\subsection{In-domain speakers and style}
\label{sec:ind}
\begin{figure}[bt!]
	\centering
	\includegraphics[width=0.9\linewidth]{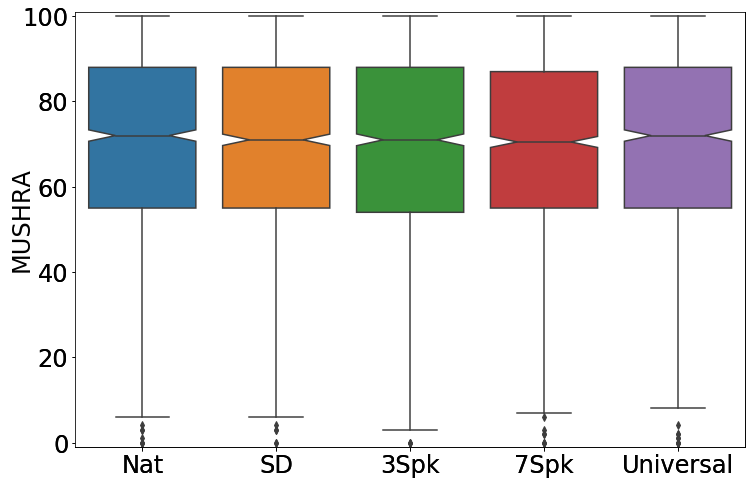}
	\vspace{-2mm}
	\caption{MUSHRA evaluation for the in-domain speakers.}
	\label{fig:ind-boxplot}
	\vspace{-5mm}
\end{figure}

\vspace{-2mm}
This evaluation considered 2 female and 1 male speaker (the ones used to train the $3Spk$ vocoder). The results in Figure \ref{fig:ind-boxplot} show that there is no significant difference in terms of evaluated naturalness when using any of the trained vocoders as long as the speakers were part of the training data. This is a strong result for the proposed universal vocoder, as it showed no degradation when compared to the highly specific $SD$ neural vocoder. Moreover, while there was a statistically significant difference between vocoded and natural naturalness scores, it was minimal (98.5\% relative MUSHRA). It must be noted that while there were inter-speaker differences, those did not affect the rank-order of the systems' ratings, so results are presented as averages.

\vspace{-4mm}
\subsection{Robustness to unseen and out-of-domain speakers}
\label{sec:ood}
\vspace{-2mm}
In this evaluation, we considered out of domain speakers for which some of the defining aspects were still part of the training corpus. That is, out of domain speakers but recorded in a studio scenario, stretching it further by considering a children audiobook scenario but from a professional 
voice talent \cite{king2016blizzard}.

In this scenario $SD$ vocoders were not available. As such, results are expectedly poor in comparison to some of the more general neural vocoders. They were included as a bottom anchor and selected by looking for the one trained with the speaker most similar to the target speaker. Similarity was measured by training a number of multi-variate Gaussian Mixture Models (GMMs) of the training data of the different vocoders and of the target speaker, then obtaining the Kullback-Leibler divergence (KLD) between the GMMs.

\vspace{-4mm}
\subsubsection{English speakers}
\begin{figure}[bt!]
	\centering
	\includegraphics[width=0.9\linewidth]{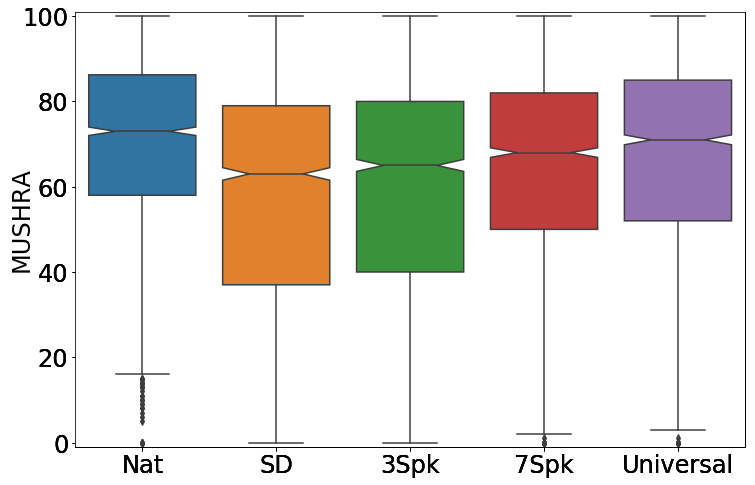}
	\vspace{-2mm}
	\caption{MUSHRA evaluation for the English, neutral, out-of-domain speakers.}
	\label{fig:ood-msp-boxplot}
	\vspace{-1mm}
\end{figure}

\vspace{-2mm}
Results (Figure \ref{fig:ood-msp-boxplot}) show that the more variety in number of training speakers the better the quality, to the point where $Univ$ is capable of providing practically the same relative MUSHRA score as for in-domain speakers (98\% vs. 98.5\%). This speaks very strongly about the generalization capabilities of such a system. Moreover, we can see how the vocoder trained with more speakers but with less training data (7Spk) is capable of providing better quality than the other two systems (SD and 3Spk), suggesting that variability is more important than quantity for generalization. 

\vspace{-4mm}
\subsubsection{Japanese speakers}
\vspace{-2mm}
We carried out an evaluation with out-of-domain Japanese speakers, which is an in-domain language, extracted from the NITech Japanese samples database \cite{zen2007hmm}. Results were similar to those in English (98\% relative MUSHRA).

\vspace{-4mm}
\subsubsection{Audiobook style speaker}
\begin{figure}[bt!]
	\centering
	\includegraphics[width=0.9\linewidth]{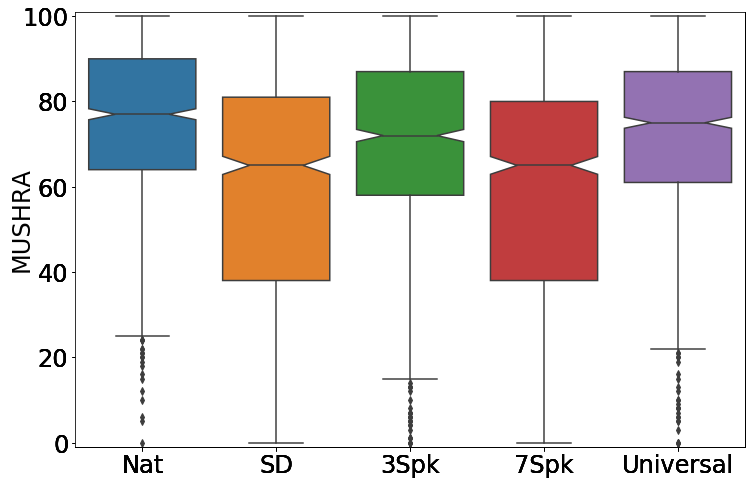}
	\vspace{-2mm}
	\caption{MUSHRA evaluation for the audiobook data.}
	\label{fig:audiobook-msp-boxplot}
	\vspace{-5mm}
\end{figure}

\vspace{-2mm}
In the case of highly expressive data, including disfluencies and onomatopoeias, (see Figure \ref{fig:audiobook-msp-boxplot}) the universal vocoder is still capable of proving steady quality, once again maintaining the relative MUSHRA scores at 98\%. Both SD and 7Spk show comparatively poor performance, while 3Spk breaks the trend. This is confirmed by the KLD between the audiobook speaker and those of the vocoders (2.64 against Univ, 5.42 against 3Spk, 14.45 against 7Spk and 14.62 against SD). All in all reinforcing the hypothesis that the dissimilarity between training and testing speakers is critical for performance.


\vspace{-4mm}
\subsection{Robustness to unseen scenarios}
\label{ssec:oos}
\vspace{-2mm}
For the additional evaluations, we did not consider all possible vocoders and restricted the exploration to a lower anchor ($SD$ systems, selected as in Section \ref{sec:ood}), an upper anchor (natural speech) and the proposed $Univ$ system.

Table \ref{tab:results} summarizes the results over the various unseen scenarios. It can be observed that the $Univ$ model significantly ($p < 0.01$) improves over the SD vocoder, with a relative MUSHRA gain varying between 15\% and 45\%. Despite this improvement, the proposed SI vocoder is not yet capable of providing a consistently high fidelity across all unseen scenarios, with a relative MUSHRA score falling down to 58\%.

\begin{table}[tb!]
	\caption{Summary of the results for the unseen scenarios. 'Rev.' stands for reverberation and 'Vocal.' for vocalizations.}
	\label{tab:results}
	\begin{tabular}{|l|c|c|c|c|c|}
		\hline
		& \textbf{SD} & \textbf{Univ} & \textbf{Nat} & \textbf{SD Rel.} & \textbf{Univ Rel.} \\ \hline
		\textbf{Breathy} & 38.4 & 61.9 & 67.6 & 56.8\% & 91.6\% \\ \hline
		\textbf{Pressed} & 30.9 & 63.4 & 70.9 & 43.5\% & 89.5\% \\ \hline
		\textbf{Noisy (N)} & 37.5 & 58.2 & 73.4 & 51.1\% & 79.4\% \\ \hline
		\textbf{Rev. (R)} & 35.5 & 56.2 & 73.6 & 48.2\% & 76.4\% \\ \hline
		\textbf{N+R.} & 23.0 & 39.7 & 68.7 & 33.5\% & 57.8\% \\ \hline
		\textbf{African} & 34.5 & 55.4 & 70.9 & 48.6\% & 78.1\% \\ \hline
		\textbf{Singing} & 41.2 & 52.3 & 72.6 & 56.8\% & 72.0\% \\ \hline
		\textbf{Vocal.} & 24.9 & 48.0 & 73.9 & 33.7\% & 64.9\% \\ \hline
	\end{tabular}
	\vspace{-5mm}
\end{table}

\vspace{-4mm}
\subsubsection{Robustness to voice quality}
\label{ssec:vqual}
\vspace{-2mm}

Results in terms of voice quality (Breathy and Pressed in Table \ref{tab:results}) appear to be relatively robust, with 91.6\% and 89.5\% relative MUSHRA respectively. This is a slight degradation compared to the normal phonation style provided with the corpus \cite{airas2007comparison} (96.3\%, not shown in Table \ref{tab:results}), but to a much lesser extent than for the SD model. The drop in relative MUSHRA compared to the clean recordings in Section \ref{sec:ood} most likely happens due to the data having been recorded at 16kHz, and due to an overall lower quality in the source material, with some clicks appearing in the end of recordings that are amplified in the re-synthesis process.

\vspace{-4mm}
\subsubsection{Robustness to signal quality}
\label{ssec:vsignal}
\vspace{-2mm}

Performance falls to about 78\% relative MUSHRA in the noisy or reverberant conditions (Noisy and Reverb. in Table \ref{tab:results} respectively), and even lower (~58\%) in a combination of both (N+R). The $Univ$ system however provides a comparable quality for noisy recordings regardless of them being in English (79\%) or in unseen African languages (78\%), which suffered from either reverberation or considerable background noise due to the poor recording conditions. 

The degradation in quality seems to be caused by distortion appearing in the re-synthesised material, as the vocoder did not seem to have learned how to properly render non-human sounds such as background noise or echo. This distortion ranges from a strong vibrato-like effect appearing in the case of reverberating samples, to distorted speech when attempting to generate the background noise.

\vspace{-4mm}
\subsubsection{Robustness to singing}
\label{ssec:vsinging}
\vspace{-2mm}

The vocoder was capable of handling singing re-synthesis (Singing in Table \ref{tab:results}) with an average performance of 72\% relative MUSHRA. A closer analysis of the results show some significant trend and differences depending on the style: clean singing (e.g. songwriter music, ballads...) performed at an average of 94.5\% relative MUSHRA, comparable to the results achievable with clean speech. Conversely, singing styles that rely on distortion (e.g. rock, pop) perform at a much lower quality (39.3\%). This correlates with the results achieved for conventional speech, suggesting that the underlying issue is voice quality rather than style. An additional observation is that tracks with multiple simultaneous voices are rendered with lower quality compared to a single voice, showcasing a limitation of the system.

\vspace{-4mm}
\subsubsection{Robustness to non-speech vocalizations}
\label{ssec:vnonspeech}
\vspace{-2mm}

The results, summarized by Vocal. in Table \ref{tab:results}, vary significantly with the kind of vocalization. While sounds of anger or achievement, represented as grunts or shouts in this dataset, perform with a poor average relative MUSHRA of 47.7\%, sounds of disgust or pleasure got an average of 77.9\%. This is probably due to the energy bursts present in the grunts and shouts, which are generated as heavily distorted sounds.

\vspace{-3mm}
\section{Discussion}
\label{sec:disc}
\vspace{-2mm}

Our experimental results in Section \ref{sec:res} have highlighted a few shortcomings to overcome. The $Univ$ system is not yet robust to noise or reverberation in the source materials, is sensitive to extreme energy bursts (shouts or grunts) and is not capable of properly generating spectrograms with multiple overlapping speakers. In these unseen scenarios, the proposed vocoder is capable of significantly outperforming a SD system (between 15\% and 45\% higher relative MUSHRA), but also introduces some distortion which substantially impairs the quality compared to that achieved in clean situations. Nonetheless, it is worth emphasizing that in studio-quality recordings, the proposed $Univ$ vocoder achieved a high fidelity of 98\% relative MUSHRA consistently across seen or unseen languages and styles. Those are promising clues showing that the generalization capabilities of the model can go way beyond simply replicating training conditions.

\vspace{-3mm}
\section{Conclusions}
\label{sec:conc}
\vspace{-2mm}

We have introduced a robust neural vocoder conditioned on mel-spectrograms, without any form of speaker encoding. The system was evaluated with an exhaustive framework, attempting to cover a very diverse range of in-domain and out-of domain scenarios.

Our results suggest that the proposed vocoder, trained on varied materials (74 speakers and 17 languages, all recorded in studio conditions) can significantly outperform speaker-dependent vocoders in clean unseen scenarios (relative MUSHRA score of 98\%). This is likely due to the variety seen during training, allowing the vocoder to generalize better to unseen scenarios, including singing, non-speech vocalizations or low-quality signals, achieving an average relative MUSHRA score of 72\%. 

Achieving a truly universal neural vocoder would allow for future work to focus on spectrogram estimation from text to any new speaker, language or style without being constrained by training a specific neural vocoder. But there is still room for improvement in terms of training data diversity and model expressiveness before we can claim the universality goal is achieved. The path towards that goal goes through understanding what training material will teach our vocoding systems to universally generalize.

\bibliographystyle{IEEEbib}
{\small \bibliography{bibliography}}\label{sec:refs}

\end{document}